\documentclass{aa} 
\usepackage{graphicx}
\usepackage{hyperref}
\usepackage[]{natbib}  
\usepackage{txfonts}
\usepackage{array}
\usepackage{slashbox}
\usepackage{colortbl}
\usepackage{multirow}
\usepackage{gensymb}
\newcommand{\re}{}
\newcommand{\ree}{}
\newcommand{\bl}{}
\usepackage{booktabs}

\begin{document}

   \title{FliPer$_{\textnormal{Class}}$: In search of solar-like pulsators among TESS targets}
   \titlerunning{FliPer$_{\textnormal{Class}}$}

   \author{L. Bugnet
          \inst{1,}\inst{2}
          \and R. A. Garc\'\i a\inst{1,}\inst{2} 
          \and S. Mathur\inst{3,}\inst{4}
          \and G. R. Davies\inst{5,}\inst{6} 
          \and O. J. Hall\inst{5,}\inst{6} 
          \and M. N. Lund\inst{6} 
          \and B. M. Rendle\inst{5,}\inst{6}
          }

\institute{IRFU, CEA, Universit\'e Paris-Saclay, F-91191 Gif-sur-Yvette, France\\
\email{lisa.bugnet@cea.fr}
\and
AIM, CEA, CNRS, Université Paris-Saclay, Université Paris Diderot, Sorbonne Paris Cité, F-91191 Gif-sur-Yvette, France
\and
Instituto de Astrof\'{\i}sica de Canarias, E-38200, La Laguna, Tenerife, Spain
\and 
Universidad de La Laguna, Dpto. de Astrof\'{\i}sica, E-38205, La Laguna, Tenerife, Spain
\and 
School of Physics and Astronomy, University of Birmingham, Edgbaston, Birmingham, B15 2TT, UK
\and 
Stellar Astrophysics Centre, Department of Physics and Astronomy, Aarhus University, Ny Munkegade 120, DK-8000 Aarhus C, Denmark
}


   \date{Received / Accepted}









\abstract{The NASA Transiting Exoplanet Survey Satellite, TESS, is about to provide full-frame images of almost the entire sky. The amount of stellar data to be analysed represents hundreds of millions stars, which is several orders of magnitude more than the number of stars observed by the Convection, Rotation and planetary Transits satellite (CoRoT), and NASA \emph{Kepler} and K2 missions. We aim at automatically classifying the newly observed stars with near real-time algorithms to better guide the subsequent detailed studies. In this paper, we present a classification algorithm built to recognise solar-like pulsators among classical pulsators. This algorithm relies on the global amount of power contained in the power spectral density (PSD), also known as the flicker in spectral power density (FliPer). Because each type of pulsating star has a characteristic background or pulsation pattern, the shape of the PSD at different frequencies can be used to characterise the type of pulsating star. The FliPer classifier (FliPer$_{\textnormal{Class}}$) uses different FliPer parameters along with the effective temperature as input parameters to feed a ML algorithm in order to automatically classify the pulsating stars observed by TESS. Using noisy TESS-simulated data from the TESS Asteroseismic Science Consortium (TASC), we classify pulsators with a $98 \%$ accuracy. Among them, solar-like pulsating stars are recognised with a $99\%$ accuracy, which is of great interest for a further seismic analysis of these stars, which are like our Sun. Similar results are obtained when we trained our classifier and applied it to 27-day \bl{subsets} of real \emph{Kepler} data. FliPer$_{\textnormal{Class}}$ is part of the large TASC classification pipeline developed by the TESS Data for Asteroseismology (T'DA) classification working group.}



  \maketitle

\section{Introduction}
Starting with the Convection, Rotation and planetary Transits satellite (CoRoT), and showing its full potential with \emph{Kepler}, asteroseismology is now the most precise way to obtain estimates of masses and radius of field stars  \citep[e.g.][]{2014A&A...569A..21L}, \ree{except for eclipsing binaries, for which spectroscopy prevails}. Asteroseismic parameters such as the frequency of maximum power $\nu_{\textnormal{max}}$ and the large frequency separation $\Delta \nu$ of the oscillation modes of solar-like pulsators \citep[i.e. with modes excited by turbulent convection,][]{1977ApJ...212..243G} are obtained from the power density spectrum using global seismic pipelines \citep[e.g.][etc.]{2009A&A...508..877M,2009CoAst.160...74H,2010A&A...511A..46M}. These global seismic parameters are key constraints for stellar evolution models: using them leads to age estimates that are much more precise than estimates obtained with other classical methods \citep[e.g.][]{2014A&A...569A..21L}.

The Transiting Exoplanet Survey Satellite (TESS), launched on 18 April  2018, conducts a photometric survey of $90\%$ of the sky during its two-year nominal operations \citep{2014SPIE.9143E..20R}. It will search for extrasolar planets that mostly orbit M-type stars. The TESS fields cover $26$ sky sectors that each cover four $24\degree$ x $24\degree$ areas from the galactic pole to nearly the ecliptic plane. 
Each field of view remains unchanged for $27$ continuous days. 
The satellite will specifically observe no fewer than 200 000 main-sequence dwarf stars, $30-100$ times brighter \bl{\citep[with an apparent magnitude lower than $\sim$ 10,][]{2018AJ....156..102S}} than those observed by the \emph{Kepler} satellite. All these conditions are suitable for seismic detections in solar-like stars, mostly in high-luminosity main-sequence (MS) and subgiant stars \cite[a detailed study of the potential asteroseismic yields of the TESS mission is given by][]{2016ApJ...830..138C}. In addition, more than $400$ million stars will be observed in the full-frame images with a $30$-minute observational cadence.\\

The first step for the large \ree{asteroseismic} survey analysis is to distinguish solar-like pulsators from all other pulsating stars. \bl{An accurate stellar classification can be computationally expensive, but efforts have been made to classify CoRoT and \textit{Kepler} targets \citep{2009A&A...506..519D,2018A&A...620A.127M}}. For example, \citet{2016ApJ...827...50M} showed three years after the end of the \emph{Kepler} main mission that more than $800$ red giants (RGs) (corresponding to about 3\%\ of the total number of observed RGs) were still misclassified as cool dwarfs \citep[see also][]{2019arXiv190300115H}. \bl{However, no public real-time automatic algorithm was developed to classify stars that wereobserved during these missions}. In view of the huge amount of data to be delivered by TESS, \ree{it would be advantageous to have an automatic method to classify solar-like stars, and even other pulsator types.}\\ 

FliPer is a method for estimating surface gravities (from 0.3 to 4.5 dex) or $\nu_{\textnormal{max}}$ of solar-type stars \citep{2018arXiv180905105B, 2017sf2a.conf...85B}. It relies on the use of the global amount of power contained in the power spectrum density (PSD) of a solar-type pulsator, which depends on its evolutionary state \citep{2011ApJ...741..119M, 2016SciA....250654K}. The method is automatic, and takes advantage of a random forest ML regressor \citep{Breiman2001} to estimate precise surface gravities. The algorithm is trained to learn how to predict $\log g$ from thousands of precise seismic estimates made with the A2Z seismic pipeline \citep{2010A&A...511A..46M}. In this way, FliPer gives estimates \bl{with a precision that  is better than can be obtained from spectroscopy alone.}\\

\begin{figure*}[!h]
 \centering  
 \includegraphics[height=11cm,clip]{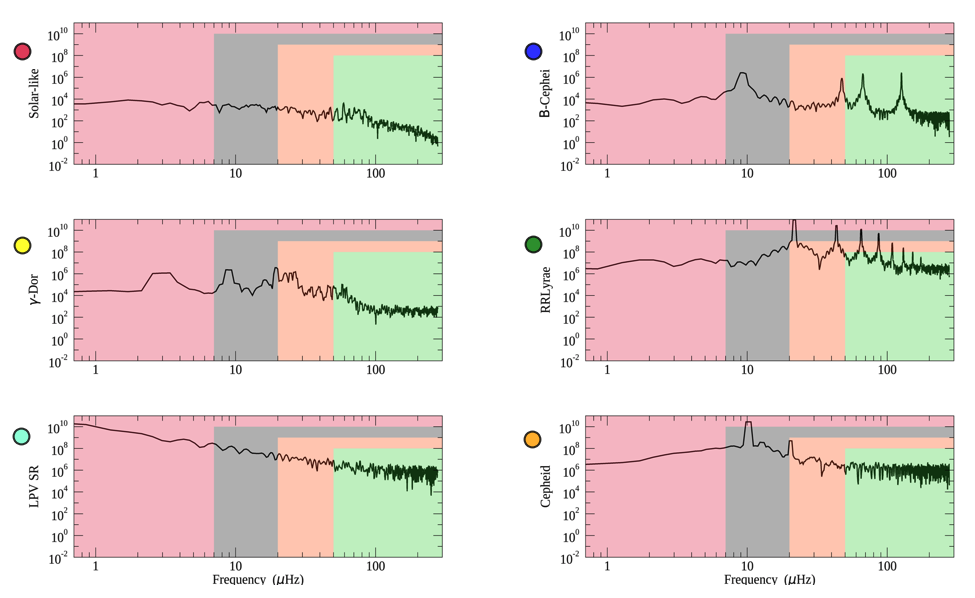}    
  \caption{PSD of six different simulated stars belonging to different classes (solar-like, $\beta$-Cephei, $\gamma$-Dor, RRLyrae, LPV, and Cepheid) as described by the $y$ -axis labels of each panel. Coloured areas (red, grey, orange, and green) represent the different ranges of frequency used for the $\textnormal{F}_{\textnormal{p},i}$ calculation (from $0.7$, $7$, $20,$ and $50$ $\mu$Hz, respectively, to the Nyquist frequency). Coloured circles represent the class identifiers used in Fig.\ref{Fig1}.}
  \label{Fig2}
\end{figure*}

\ree{Machine-learning methods such as neural networks \citep[e.g.][]{2005ChJAA...5..203B}, algorithms based on decision trees \citep[e.g.][]{2017A&A...605A.123P, 2018arXiv180405245V}, or AdaBoost \citep[e.g.][]{2018arXiv180405051V} already give good results for characterising the stars. For instance, \cite{2018MNRAS.tmp..471H} showed that they were able to distinguish core helium-burning clump stars from hydrogen shell-burning RG  stars using a convolutional neural network.} In our study, \bl{we use FliPer parameters to classify solar-like pulsators from among all pulsating stars: instead of using a regressor \citep[see][]{2018arXiv180905105B} to estimate physical parameters, we use a classifier algorithm trained with the FliPer parameters and the effective temperature of each star.} After describing the data in Section~2, we explain  in Section~3 how the FliPer$_{\textnormal{Class}}$ algorithm uses FliPer parameters ($\textnormal{F}_{\textnormal{p},i}$) along with the effective temperature to distinguish between the different pulsator types. Then we present results from the classification of TESS-simulated data and of a known sample of \emph{Kepler} main mission data. 

\section{Data preparation}

In order to test the algorithms, the T'DA working group simulated datasets of TESS observations\footnote{Datasets can be downloaded after registration on the TESS Asteroseismic Science Operations Center (TASOC) website at \url{https://tasoc.dk/wg0/SimData}} \citep{2017EPJWC.16001005L}. We used $10,812$ simulated stars that can be studied with a stellar signal alone (designated as ``clean'' data), with additional white noise (``noisy'' data), or with both additional white noise and instrumental systematics (``sysnoisy'' data). Because systematics can be corrected \citep[using methods similar to those applied to the K2 data,][]{2016MNRAS.459.2408A}, we chose to focus our study on the ``noisy'' dataset. The sample is described in Tab.~\ref{Tab1}. Part of the $\gamma$-Doradus sample is constituted of $\gamma$-Doradus and $\delta$-Scuti hybrid stars.\\ 

 \begin{table}[t]
\caption{Composition of the samples from T'DA simulated dataset and real \emph{Kepler} data.
\label{param}}             
\centering
\begin{tabular}{@{} l|*{2}{>{$}c<{$}} @{}}   
   \toprule
Type of star & \multicolumn{1}{c@{}}{TESS$_{simulated}$} & \multicolumn{1}{c@{}}{\emph{Kepler}}\\   
\hline
&&\\
Solar-like (SL) &3668 & 802 \\
Subdwarf B (sdBV) & 129& 8\\
$\beta$-Cephei ($\beta$-Cep) & 298 & 5\\
Slowly pulsating B-type (SPB) & 1846 & 26\\
$\delta$-Scuti &115 & 358\\
$\gamma$-Doradus ($\gamma$-Dor) & 1569 & 202\\
rapidly oscillating Ap (roAp) & 287 & 3\\
RRLyrae & 646 & 36\\
Long-period variable (LPV) & 965 & 0\\
Cepheid & 1289 & 2\\
\end{tabular}

\end{table}
To determine the reliability of our method on real data, we also used power spectrum densities of a sample of $1,442$ \emph{Kepler} targets observed in the long-cadence observation mode (corresponding to an acquisition every 30 min) for which we know the classification. Tab.~\ref{Tab1} displays the number of stars in the Kepler sample belonging to each classification \citep{2018OAst...27..157R, 2012AJ....143..101M, 2019MNRAS.482.1757L,2011MNRAS.410..517B, 2013MNRAS.436.1415B, 2014psce.conf..315S,2015MNRAS.452.3334S,2017ApJS..233...23S}. Long-period variability stars are not represented in the \emph{Kepler} sample because they can be easily classified by using the effective temperature alone (the FliPer$_{\textnormal{Class}}$ is not required for these stars). The \emph{Kepler} light curves \citep[calibrated following][]{2011MNRAS.414L...6G} considered in this work were observed for approximately four years. This results in a much higher frequency resolution in the PSD than what is expected for most TESS targets, which are observed for only 27 days. To test our method on data that are representative of the first sector of TESS data, we computed the PSD of each star based on randomly extracted 27-day periods of time from the full \emph{Kepler} time series. \ree{We also used the effective temperatures from \cite{0067-0049-229-2-30} for the sample of \textit{Kepler} stars.}

\section{Fliper$_{\textnormal{Class}}$: a tool for classifying pulsating stars}

FliPer \citep{2018arXiv180905105B} is a method for estimating the surface gravity of solar-like pulsating stars based on the measure of the amount of power in their PSD. For solar-type pulsators, the PSD is dominated by the power of the convective background, stellar oscillation modes, and the rotation period signals. All these effects vary when the star evolves from the MS to the red giant branch (RGB). FliPer thus gives constraints on the evolutionary stage of the solar-like pulsator. We define the FliPer metric as 

\begin{equation}
   \textnormal{F}_{\textnormal{p}} = \overline{\textnormal{PSD}} - \textnormal{P}_\textnormal{n} ,
   \label{powvar}
\end{equation}
     
where $\overline{\textnormal{PSD}}$ represents the averaged value of the PSD from a given frequency to the Nyquist frequency, and $\textnormal{P}_\textnormal{n}$ is the photon noise \citep[see][for more information]{2017sf2a.conf...85B}. \\

\subsection{FliPer parameters: $\textnormal{F}_{\textnormal{p},i}$}
For each star we calculated different FliPer parameters, $\textnormal{F}_{\textnormal{p},i}$, \ree{as the FliPer metric} starting from different \ree{lower} frequency boundaries ($i \in [0.7, 7, 20, 50]$ $ \mu$Hz) in the calculation of $\overline{\textnormal{PSD}}$. The four different frequency domains used for the $\textnormal{F}_{\textnormal{p},i}$ calculation are represented by the coloured area in Fig.~\ref{Fig2}. By combining these different $\textnormal{F}_{\textnormal{p},i}$, we extracted information from different regions of the PSD of the star. A previous study \citep[see][]{2017sf2a.conf...85B} indicated that the two $\textnormal{F}_{\textnormal{p},0.7}$ and $\textnormal{F}_{\textnormal{p},7}$ parameters are easily dominated by rotation peaks for MS stars, but are perfectly suitable to take the power of the modes for high-luminosity giants into account. The other parameters, $\textnormal{F}_{\textnormal{p},20}$ and $\textnormal{F}_{\textnormal{p},50}$, allow precise estimates for MS stars but they do not take the mode power in high-luminosity RGs into account. FliPer gives great results when MS stars are distinguished from RGs by estimating their surface gravity, as discussed in \cite{2018arXiv180905105B}. By combining the different $\textnormal{F}_{\textnormal{p},i}$ for all stellar types, we attempted to classify not only solar-like stars, but all pulsator types.\\

Each pulsator type has a typical amount of power associated for a given frequency range in the PSD. Figure~\ref{Fig2} shows the TESS-simulated PSD for six different pulsator types. First, we observe that each type of star presents a characteristic \bl{signature in the PSD}. 

By calculating $\textnormal{F}_{\textnormal{p},0.7}$ (red areas on Fig.~\ref{Fig2}), it is easy to distinguish a solar-like star from a long-period variable (LPV) because their granulation power differs by several orders of magnitude. However, it is harder to distinguish a Cepheid from a RRLyrae using only $\textnormal{F}_{\textnormal{p},0.7}$ because they both present a PSD background with the same order of magnitude. With a higher frequency boundary such as $50$ $\mu$Hz for the $\textnormal{F}_{\textnormal{p},i}$ calculation, we can distinguish a Cepheid well from a RRLyrae. However, by simultaneously using the different $\textnormal{F}_{\textnormal{p},i}$ , it is possible to distinguish the different types of stellar pulsators.

\newcolumntype{L}[1]{>{}p{#1}}
\newcolumntype{C}[1]{>{\centering\arraybackslash}p{#1}}

\begin{figure*}[bt]
 \centering
  \includegraphics[height=7cm,clip]{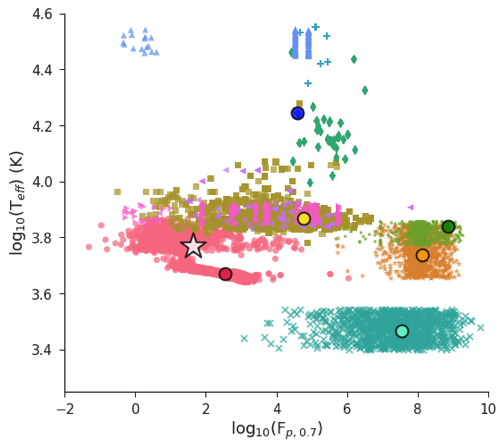}%
   \includegraphics[height=7cm,clip]{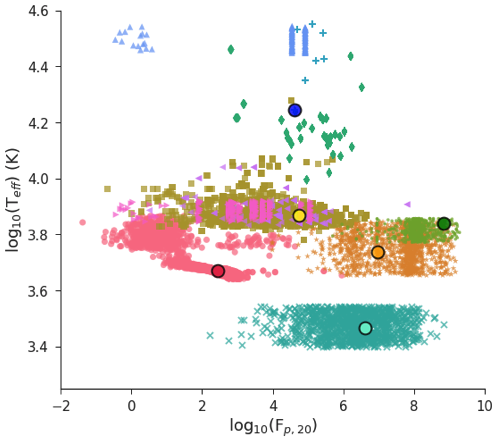} 
   \includegraphics[height=5.5cm,clip]{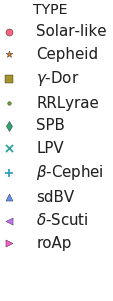}  

  \caption{{\bf Left panel}: Representation of the total sample of simulated TESS stars in the $\log$(\textsl{T}$_{\rm eff}$) vs. $\log(\textnormal{F}_{\textnormal{p},0.7})$ diagram. Each stellar type is associated with a unique colour \bl{and symbol} reported at the \bl{side} legend. In addition, the positions of the stars shown in Fig.~\ref{Fig2} are added to the diagram and are represented with circles. The white star represents the position in the diagram of the TESS target TIC 261136679. {\bf Right panel}: Same as the left panel, but for $\textnormal{F}_{\textnormal{p},20}$.}
  \label{Fig1}
\end{figure*}

 As previously discussed, the type of variability shown by the star affects the range of values that it can have for each $\textnormal{F}_{\textnormal{p},i}$. Figure~\ref{Fig1} represents the total sample of TESS-simulated data in the $\log$(\textsl{T}$_{\rm eff}$) versus $\textnormal{F}_{\textnormal{p},i}$ diagram for $i=0.7$ $\mu$Hz (left panel) and $i=20$ $\mu$Hz values (right panel). In addition, the stars shown in Fig.~\ref{Fig2} are represented in the diagrams with stars with the same colour code as in Fig.~\ref{Fig1}. \bl{We also represent the first planet star host observed by TESS in the $\pi$-Mensae system with a star \citep{2018arXiv180905967H, 2018arXiv180907573G}}. \re{This star is properly classified as a solar-like pulsator based on its  FliPer values, as shown in Fig.~\ref{Fig1}.}\\

Using the TESS simulated dataset, we note that each type of star covers a given region of the \textsl{T}$_{\rm eff}$ versus $\textnormal{F}_{\textnormal{p},i}$ diagrams. \bl{This} means for instance that using only one $\textnormal{F}_{\textnormal{p},i}$, solar-like pulsators are already well separated from Cepheids and RRLyrae. However, we extended the analysis of Fig.~\ref{Fig2} and show with Fig.~\ref{Fig1} that using only one $\textnormal{F}_{\textnormal{p},i}$ does not allow us to clearly distinguish between Cepheids and RRLyrae. In addition, we  observe by comparing the two panels of Fig.~\ref{Fig1} that the area corresponding to a given type of star changes when a different $\textnormal{F}_{\textnormal{p},i}$ is used: each pulsator type evolves differently in the diagram when we modify the starting frequency of the $\textnormal{F}_{\textnormal{p},i \in [0.7, 7, 20, 50]}$ calculation (when we switch from the left to the right panel in Fig.~\ref{Fig1}). We therefore expect to be able to separate RRLyrae from Cepheids by comparing their different $\textnormal{F}_{\textnormal{p},i}$ ($i \in [0.7, 7, 20, 50]$ $ \mu$Hz).\\

\begin{table*}[t]
\centering
\caption{Confusion matrix of the TESS-simulated data test sample. Values represent the number of stars, and italic numbers in parentheses represent the percentage accuracy for the class. The colour code is the same as in Fig.~\ref{Fig1} and is normalised for each row by the total number of stars in each true class. Numbers that do not belong to the diagonal represent classification errors by FliPer$_{\textnormal{Class}}$.
\label{Tab1}}             
\centering
\begin{tabular}{| L{3.3cm}|C{0.9cm}C{0.9cm}C{0.9cm}C{0.9cm}C{0.9cm}C{0.9cm}C{0.9cm}C{0.9cm}C{0.9cm}C{0.9cm}|} 
\hline
\backslashbox{True}{Predicted\,\,\,\,}&S-l&sdBV&$\beta$&SPB&$\delta$&$\gamma$&roAp&RR&LPV&Ce\\
\hline
\multirow{2}{2cm}{Solar-like} & \cellcolor[cmyk]{0,1,0.4,0.01}713 {(\it 99.9)}&  \cellcolor[cmyk]{0,0,0,0} &  \cellcolor[cmyk]{0,0,0,0}  & \cellcolor[cmyk]{0,0,0,0}  & \cellcolor[cmyk]{0,0,0,0}  &  &  \cellcolor[cmyk]{0,0,0,0}  & \cellcolor[cmyk]{0,0.1,0.04,0.01}\textcolor{black}{1} &  \cellcolor[cmyk]{0,0,0,0} &  \cellcolor[cmyk]{0,0,0,0}\rule[-7pt]{0pt}{20pt}\\

\multirow{2}{2cm}{sdBV} & \cellcolor[cmyk]{0,0,0,0} & \cellcolor[cmyk]{1,0,0,0.5}33 {(\it 100)} &  \cellcolor[cmyk]{0,0,0,0}  & \cellcolor[cmyk]{0,0,0,0} &  \cellcolor[cmyk]{0,0,0,0} &   & \cellcolor[cmyk]{0,0,0,0} & \cellcolor[cmyk]{0,0,0,0} &  \cellcolor[cmyk]{0,0,0,0}  & \cellcolor[cmyk]{0,0,0,0}\rule[-7pt]{0pt}{20pt}\\

\multirow{2}{2cm}{$\beta$-Cep} &  \cellcolor[cmyk]{0,0,0,0} &  \cellcolor[cmyk]{0,0,0,0} & \cellcolor[cmyk]{1,0,0.3,0.3}52 {(\it 100)}  & \cellcolor[cmyk]{0,0,0,0}  & \cellcolor[cmyk]{0,0,0,0}  & \cellcolor[cmyk]{0,0,0,0}  & \cellcolor[cmyk]{0,0,0,0}  & \cellcolor[cmyk]{0,0,0,0}  & \cellcolor[cmyk]{0,0,0,0}  & \cellcolor[cmyk]{0,0,0,0}\rule[-7pt]{0pt}{20pt}\\

{SPB} &\cellcolor[cmyk]{0,0,0,0} &  \cellcolor[cmyk]{0,0,0,0} &  \cellcolor[cmyk]{0,0,0,0} &\cellcolor[cmyk]{1,0,0.9,0}360 {(\it 100)}  & \cellcolor[cmyk]{0,0,0,0}  & \cellcolor[cmyk]{0,0,0,0}  & \cellcolor[cmyk]{0,0,0,0}  & \cellcolor[cmyk]{0,0,0,0} &  \cellcolor[cmyk]{0,0,0,0}   &\cellcolor[cmyk]{0,0,0,0}\rule[-7pt]{0pt}{20pt}\\

{$\delta$-Scuti}& \cellcolor[cmyk]{0,0,0,0}  & \cellcolor[cmyk]{0,0,0,0}  & \cellcolor[cmyk]{0,0,0,0}  & \cellcolor[cmyk]{0,0,0,0} &  \cellcolor[cmyk]{0.4,0.8,0.1,0}18 {(\it 64.3)}& \cellcolor[cmyk]{0.2,0.4,0.05,0}\textcolor{black}{9}  & \cellcolor[cmyk]{0.04,0.08,0.01,0}{1} &  \cellcolor[cmyk]{0,0,0,0}  & \cellcolor[cmyk]{0,0,0,0}  & \cellcolor[cmyk]{0,0,0,0}\rule[-7pt]{0pt}{20pt}\\

{$\gamma$-Dor}& \cellcolor[cmyk]{0.06,0,0.2,0}\textcolor{black}{ 7}  & \cellcolor[cmyk]{0,0,0,0}  & \cellcolor[cmyk]{0,0,0,0}   &\cellcolor[cmyk]{0.05,0,0.12,0}\textcolor{black}{ 3}  & \cellcolor[cmyk]{0.03,0,0.1,0}\textcolor{black}{ 2} &\cellcolor[cmyk]{0.3,0,1,0.2}320 {(\it 96.1)} & \cellcolor[cmyk]{0.03,0,0.1,0}\textcolor{black}{ 1}  & \cellcolor[cmyk]{0,0,0,0} &  \cellcolor[cmyk]{0,0,0,0} &  \cellcolor[cmyk]{0,0,0,0}\rule[-7pt]{0pt}{20pt}\\

roAp & \cellcolor[cmyk]{0,0.1,0,0.05}\textcolor{black}{ 1} &  \cellcolor[cmyk]{0,0,0,0} &  \cellcolor[cmyk]{0,0,0,0}  & \cellcolor[cmyk]{0,0,0,0}  & \cellcolor[cmyk]{0,0,0,0}  &  &\cellcolor[cmyk]{0,1,0,0.1}50 {(\it 98.1)} &  \cellcolor[cmyk]{0,0,0,0} &  \cellcolor[cmyk]{0,0,0,0} &  \cellcolor[cmyk]{0,0,0,0}\rule[-7pt]{0pt}{20pt}\\

RRLyrae & \cellcolor[cmyk]{0,0,0,0} &  \cellcolor[cmyk]{0,0,0,0}  & \cellcolor[cmyk]{0,0,0,0}  & \cellcolor[cmyk]{0,0,0,0} &  \cellcolor[cmyk]{0,0,0,0}  & \cellcolor[cmyk]{0,0,0,0}   &\cellcolor[cmyk]{0,0,0,0} &\cellcolor[cmyk]{0.6,0,1,0.02}115 {(\it 98.3)} &  \cellcolor[cmyk]{0,0,0,0} & \cellcolor[cmyk]{0.1,0,0.1,0.002}\textcolor{black}{ 2}\rule[-7pt]{0pt}{20pt}\\
\arrayrulecolor{black}
LPV &  \cellcolor[cmyk]{0,0,0,0}  &  \cellcolor[cmyk]{0,0,0,0} &  \cellcolor[cmyk]{0,0,0,0}   &\cellcolor[cmyk]{0,0,0,0}  & \cellcolor[cmyk]{0,0,0,0}  & \cellcolor[cmyk]{0,0,0,0} &  \cellcolor[cmyk]{0,0,0,0} &  \cellcolor[cmyk]{0,0,0,0}& \cellcolor[cmyk]{1,0,0.5,0}214 {(\it 100)} &  \cellcolor[cmyk]{0,0,0,0}\rule[-7pt]{0pt}{20pt}\\

Cepheid & \cellcolor[cmyk]{0,0,0,0} &  \cellcolor[cmyk]{0,0,0,0} &  \cellcolor[cmyk]{0,0,0,0}  & \cellcolor[cmyk]{0,0,0,0}  & \cellcolor[cmyk]{0,0,0,0}  & \cellcolor[cmyk]{0,0.03,0.1,0}\textcolor{black}{ 1} &  \cellcolor[cmyk]{0,0,0,0} & \cellcolor[cmyk]{0,0.07,0.2,0}\textcolor{black}{ 6}   &\cellcolor[cmyk]{0,0,0,0}\rule[-7pt]{0pt}{20pt}&\cellcolor[cmyk]{0,0.6,1,0} 256 {(\it 96.6)}\\
\hline
\end{tabular}
\end{table*}

\subsection{ FliPer$_{\textnormal{Class}}$ classification algorithm}

In the previous section we explained that stars can be manually classified according to their $\textnormal{F}_{\textnormal{p},i\in [0.7, 7, 20, 50]}$. In view of the amount of TESS data to be released, the classification of each individual pulsator has to be automatic. A random forest classifier \citep{Breiman2001} is a supervised machine-learning (ML) algorithm that classifies data from a given set of input parameters (see Appendix \ref{random} for more details about the classifier). 
Random forest algorithms have been proven to be efficient in distinguishing between MS stars and RGs \citep{2018arXiv180905105B} when $\textnormal{F}_{\textnormal{p},i}$ ($i \in [0.7, 7, 20, 50]$ $\mu$Hz) is used as input parameters.\\

\begin{figure}[b!]
 \centering
 \includegraphics[width=0.5\textwidth,clip]{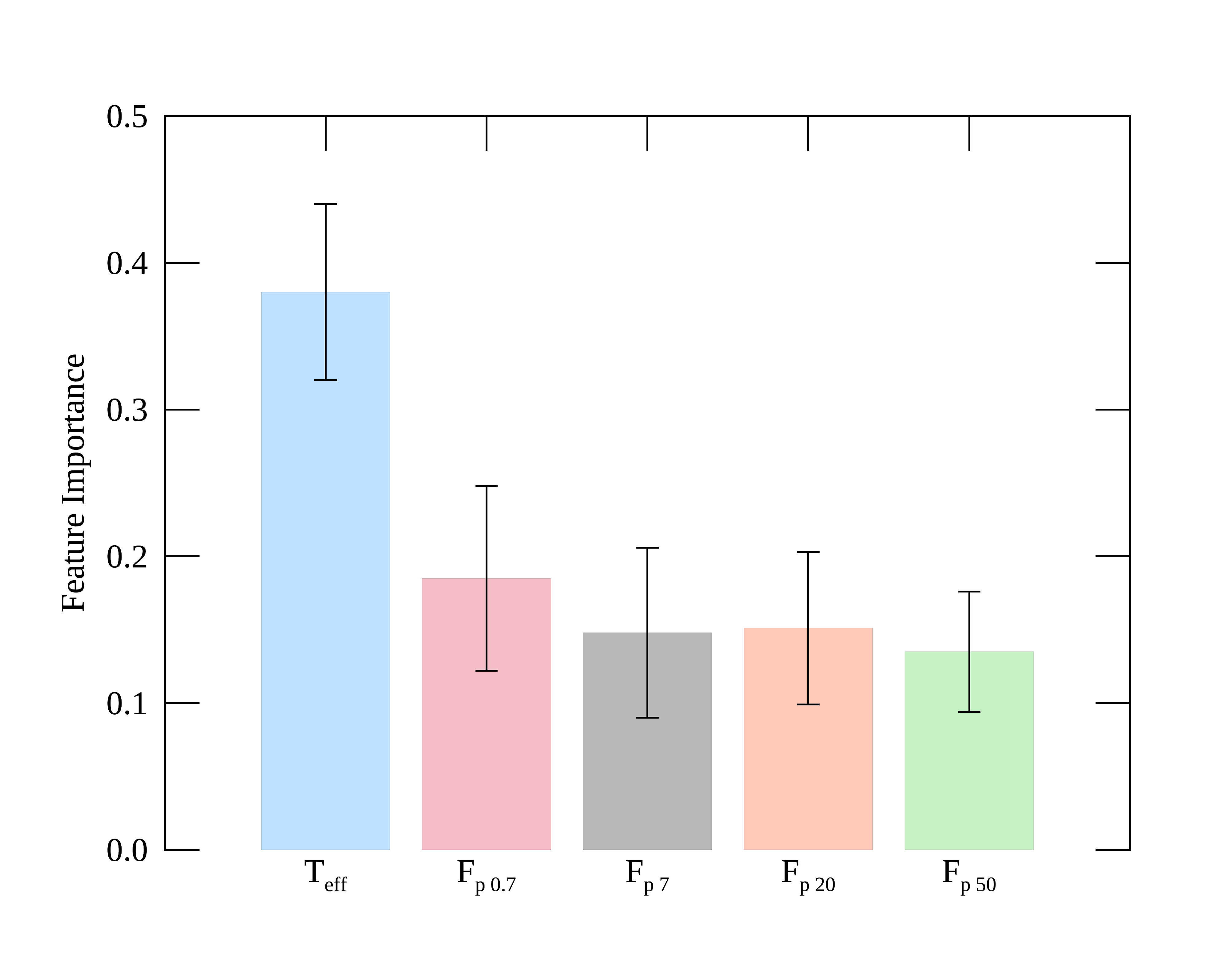}
  \caption{Significance of the different input parameters on the training process \re{based on the TESS-simulated dataset} \ree{along with their uncertainties}.}
  \label{Fig3}
\end{figure}

We classified the pulsators using the "RandomForestClassifier" function from the "sklearn.ensemble" Python library \citep{scikit-learn}. We split the simulated dataset into two random samples. The training set contains $80\%$ of the total number of stars, while the test sample contains the remaining $20\%$ of the stars. The same method was applied to the \textit{Kepler} set. The supervised classifier FliPer$_{\textnormal{Class}}$ was trained on the training dataset to learn how to predict the output classification using $\textnormal{F}_{\textnormal{p},i}$ ($i \in [0.7, 7, 20, 50]$ $\mu$Hz) and \textsl{T}${_{\rm eff}}$ as input parameters for each star.
\ree{The maximum number of features considered at each split point is $p=\sqrt{m}$ because we consider $m=5$ input parameters (see Appendix~\ref{random} for details about the classifier).} The previously trained algorithm, along with the code to use it, can be downloaded from GitHub\footnote{\url{https://github.com/lbugnet/FLIPER_CLASS}}. Each parameter has a different effect on the training process, which is represented in Fig.~\ref{Fig3} for the TESS-simulated dataset. Feature importance is the number of times a feature is used to split a node normalised by the total number of nodes. Uncertainties are calculated by taking the standard deviation of each feature importance from the individual trees. \ree{The effective temperature has the highest weight in classifying the type of stars. However, all input parameters are useful regarding the importance of the other $\textnormal{F}_{\textnormal{p},i}$ parameters.} This shows that $\textnormal{F}_{\textnormal{p},i}$ parameters, coupled with \textsl{T}$_{\rm eff}$, are suitable parameters for classifying stars. Similar results are obtained when the \textit{Kepler} training sample was used.

\section{\bl{Classification of TESS-simulated data}}

We obtained an out-of-bag (OOB) error of the training on TESS-simulated data of about 0.011. This number gives estimates of the error rate of the classifier when $n_t=200$ trees are used by classifying a sub-sample of stars that were not used in the building of the last learner. The OOB error can be biased depending on the hyperparameters of the algorithm \citep[number of trees ($n_t$), number of features considered at each split point ($m$), etc,][]{ooberror}. \bl{This study thus provides another estimate of the classification accuracy using the TESS test sample to examine the performance of the trained algorithm.}

The $\sim 2,000$ stars that belong to the TESS test sample were automatically classified amongst the classes reported in Tab. \ref{Tab1} by FliPer$_{\textnormal{Class}}$ trained on the training sample. The results are represented in Tab.~\ref{Tab1}: the numbers in each row represent for a given pulsator class the number of stars that were classified in each output class by FliPer$_{\textnormal{Class}}$. The higher the value on the diagonal (a high value corresponds to a dark-coloured cell), the more accurate the algorithm for the corresponding class. From this table, we first conclude that $\sim 98\%$ of stars in the test set are well characterised by the algorithm.\\

Then, considering specifically the misclassified stars, we show that most classification errors concern classical pulsators. In particular, $\delta$-Scutis are classified as $\gamma$-Doradus, which can be explained by \re{the fact that there are not enough $\delta$-Scuti stars in the sample for the algorithm to learn how to recognise this type of star, and also because the $\gamma$-Doradus training sample contains some $\gamma$-Doradus and $\delta$-Scuti hybrid stars.} This misclassification problem should be solved with real TESS data as the training of the algorithm will be made on a larger set of stars \bl{that belong to each of the categories, including new hybrid categories, which will allow us to separate similar classes such as $\delta$-Scuti and $\gamma$-Dor well.} From studying the spectra, we note that most misclassified solar-like and roAp stars show nearly flat power spectra, except at very low frequency. It is already known from \cite{2018arXiv180905105B} that FliPer, and thus FliPer$_{\textnormal{Class}}$, is not efficient for this type of noise-dominated spectrum because it compares the global amount of power in the power spectrum with the power at high frequency (representing the photon noise).\\

\begin{table*}[h]
\centering
\caption{Confusion matrix of the \emph{Kepler} data test sample \bl{when $\textnormal{F}_{\textnormal{p},i}$ ($i \in [0.7, 7, 20, 50]$ $\mu$Hz) and \textsl{T}$_{\rm eff}$ are used as input parameters for each star}. Values represent the number of stars, and italic numbers in parentheses represent the percentage accuracy for the class. The colour code is normalised for each row by the total number of stars in each true class. Numbers that do not belong to the diagonal represent classification errors by FliPer$_{\textnormal{Class}}$.\label{Tab2}}             
\centering
\begin{tabular}{| L{3.3cm}|C{2.2cm}C{2.2cm}C{2.2cm}C{2.2cm}C{2.2cm}|} 
\hline

\backslashbox{True}{Predicted\,\,\,\,}&S-l&RR/Cep&$\gamma$-Dor&$\delta$/roAp/sdBV&SPB/$\beta$\\
\hline
Solar-like & \cellcolor[cmyk]{0,1,0.4,0.01}{161} {(\it 100)}&  \cellcolor[cmyk]{0,0,0,0} &  \cellcolor[cmyk]{0,0,0,0}  & \cellcolor[cmyk]{0,0,0,0}  & \cellcolor[cmyk]{0,0,0,0}\rule[-7pt]{0pt}{20pt}\\

RRLyrae/Cepheid & \cellcolor[cmyk]{0,0,0,0} & \cellcolor[cmyk]{0.6,0,1,0.02}9 {(\it 100)}&    &   &  \cellcolor[cmyk]{0,0,0,0}\rule[-7pt]{0pt}{20pt}\\

$\gamma$-Dor &  \cellcolor[cmyk]{0,0,0,0} &  \cellcolor[cmyk]{0,0,0,0} & \cellcolor[cmyk]{0.3,0,1,0.2}41 {(\it 100)}  & \cellcolor[cmyk]{0,0,0,0}  & \cellcolor[cmyk]{0,0,0,0}\rule[-7pt]{0pt}{20pt}\\

$\delta$-Scuti/RoAp/sdBV & \cellcolor[cmyk]{0,0,0,0} &  \cellcolor[cmyk]{0,0,0,0} &  \cellcolor[cmyk]{0.04,0.08,0.01,0}\textcolor{black}{ 1}  & \cellcolor[cmyk]{0.4,0.8,0.1,0}73 {(\it 97.3)} & \cellcolor[cmyk]{0.04,0.08,0.01,0}\textcolor{black}{ 1}\rule[-7pt]{0pt}{20pt}\\

SPB/$\beta$-Cephei & \cellcolor[cmyk]{0,0,0,0} &  \cellcolor[cmyk]{0,0,0,0}  & \rule[-7pt]{0pt}{20pt}  & \cellcolor[cmyk]{0.2,0,0.18,0}\textcolor{black}{ 1} &  \cellcolor[cmyk]{1,0,0.9,0}6 {(\it 85.7)}\\
\arrayrulecolor{black}\hline
\end{tabular}
\end{table*}

\begin{table*}[h]
\centering
\caption{\bl{Confusion matrix of the \emph{Kepler} data test sample when $\textnormal{F}_{\textnormal{p},i}$ ($i \in [0.7, 7, 20, 50]$ $\mu$Hz) alone is used as input parameter for each star. Values represent the number of stars, and italic numbers in parentheses represent the percentage accuracy for the class. The colour code is normalised for each row by the total number of stars in each true class. Numbers that do not belong to the diagonal represent classification errors.}\label{Tab3}}             
\centering
\begin{tabular}{| L{3.3cm}|C{2.2cm}C{2.2cm}C{2.2cm}C{2.2cm}C{2.2cm}|} 
\hline

\backslashbox{True}{Predicted\,\,\,\,}&S-l&RR/Cep&$\gamma$-Dor&$\delta$/roAp/sdBV&SPB/$\beta$\\
\hline
Solar-like & \cellcolor[cmyk]{0,1,0.4,0.01}{161} {(\it 100)}&  \cellcolor[cmyk]{0,0,0,0} &  \cellcolor[cmyk]{0,0,0,0}  & \cellcolor[cmyk]{0,0,0,0}{}  & \cellcolor[cmyk]{0,0,0,0}\rule[-7pt]{0pt}{20pt}\\

RRLyrae/Cepheid & \cellcolor[cmyk]{0,0,0,0} & \cellcolor[cmyk]{0.6,0,1,0.02}8 {(\it 88.6)}&    &  \cellcolor[cmyk]{0.2,0,0.3,0.02}1 &  \cellcolor[cmyk]{0,0,0,0}\rule[-7pt]{0pt}{20pt}\\

$\gamma$-Dor &  \cellcolor[cmyk]{0,0,0,0} &  \cellcolor[cmyk]{0,0,0,0} & \cellcolor[cmyk]{0.3,0,1,0.2}41 {(\it 99.4)}  & \cellcolor[cmyk]{0,0,0,0}  & \cellcolor[cmyk]{0,0,0,0}\rule[-7pt]{0pt}{20pt}\\

$\delta$-Scuti/RoAp/sdBV & \cellcolor[cmyk]{0.08,0.16,0.02,0}1 &  \cellcolor[cmyk]{0,0,0,0} &  \cellcolor[cmyk]{0,0,0,0}\textcolor{black}{ }  & \cellcolor[cmyk]{0.4,0.8,0.1,0}73 {(\it 93.8)} & \cellcolor[cmyk]{0.04,0.08,0.01,0}\textcolor{black}{ 1}\rule[-7pt]{0pt}{20pt}\\

SPB/$\beta$-Cephei & \cellcolor[cmyk]{1,0,0.9,0}5 &  \cellcolor[cmyk]{0,0,0,0}  & \rule[-7pt]{0pt}{20pt} \cellcolor[cmyk]{0.2,0,0.18,0}1 & \cellcolor[cmyk]{0.2,0,0.18,0}\textcolor{black}{ 1} &  \cellcolor[cmyk]{0,0,0,0}0 {(\it 0)}\\
\arrayrulecolor{black}\hline
\end{tabular}
\end{table*}

\section{\re{Classification based on 27-day segments of real \emph{Kepler} data}}
\re{
To estimate the accuracy of the method on real data, we trained (tested) the algorithm on $80\%$ ($20\%$) of the global set of \textit{Kepler} data. As the number of some types of classical pulsators (such as sdBV, $\beta$-Cep, RoAp, and Cepheid stars) observed by the \emph{Kepler} main mission is very small (see Tab.~\ref{param}), it is too ambitious to train and test the algorithm to recognise all different types of stellar pulsators. To avoid misclassification due to the lack of stars in the \emph{Kepler} catalogue, we chose to group several pulsators into categories dependent upon their position in the Hertzsprung-Russell diagram: 
\begin{itemize}
\item{}$\delta$-Scuti, RoAp, and sdBV stars have a low luminosity ($10\textnormal{L}_\odot<\textnormal{L}<100\textnormal{L}_\odot$).
\item{}$\beta$-Cep and SPB stars have a high luminosity ($100\textnormal{L}_\odot<\textnormal{L}<100,000\textnormal{L}_\odot$) and high effective temperatures (4<$\log_{10}$(\textsl{T}$_{\rm eff}$)<4.5).
\item{}Cepheids and RRLyrae have a high luminosity ($30\textnormal{L}_\odot<\textnormal{L}<100,000\textnormal{L}_\odot$) and low effective temperature (3.6<$\log_{10}$(\textsl{T}$_{\rm eff}$)<3.9).
\end{itemize}
We then considered the five different classes reported in Tab.~\ref{Tab2}, which represents the confusion matrix for stars in the \emph{Kepler} test set. As in Tab. ~\ref{Tab1}, values in each row represent for a given class the number of stars classified in each output class by the FliPer$_{\textnormal{Class}}$. The accuracy of the classification of the \emph{Kepler} test sample is approximately $99\%$. We point out that all solar-like stars are correctly classified by the algorithm (which we recall was our main goal). Most misclassifications concern classical pulsators, with a low corresponding number of stars in the training set, which means that the training was probably more difficult for these types of stars. This problem should be solved by training the algorithm on a much larger set of TESS observations. }

\subsection{\bl{Effect of the effective temperature for the classification}}

\bl{A distribution of input parameter importance very similar to that shown in Fig.~\ref{Fig3} was obtained when we trained on the 0\textit{ Kepler} sample. The effective temperature thus seems to play a much larger role in the classification process than the different FliPer parameters. We decided to show classification results when the effective temperature was removed from the input parameters in order to explain to which extent the effective temperature is needed for the classification. \\}

When the classifier is tested and trained on \textit{Kepler} data and the effective temperature is not used as an input parameter, the classification only depends on the FliPer parameters. With this configuration, we obtain a $96\%$ accuracy on the classification of the test set. Solar-like stars are still very well classified, and most errors concern the SPB/$\beta$-Cephei class (see Tab.~\ref{Tab3}). Indeed Fig~\ref{Fig2} shows that $\textnormal{F}_{\textnormal{p},i}$ ($i \in [0.7, 7, 20, 50]$ $\mu$Hz) values for SPB/$\beta$-Cephei are quite similar to those of solar-like stars and of $\delta$-Scuti, RoAp, and sdBV.\\

\bl{With this study, we point out that the FliPer parameters alone as input to the algorithm are enough to recognise all solar-like stars. However, adding physical parameters (such as \textsl{T}$_{\rm eff}$) to the classifier allows FliPer$_{\textnormal{Class}}$ to perform well for all pulsators, and also to avoid false detection of solar-like stars, as shown in Tab.~\ref{Tab2}.}

\subsection{\bl{Taking uncertainties on input parameters into account}}

\bl{
In order to test the robustness of the classifier regarding uncertainties on the input parameters, we tested the algorithm on the \textit{Kepler} dataset with modified input parameter values.
 Uncertainties on FliPer arise from the photon noise in the spectra (following a chi-squared distribution with two degrees of freedom). Hence, the uncertainty on the FliPer \citep[see][for more details]{2018arXiv180905105B} parameters can be explicitly written as 
\begin{equation}
\bl{\delta \textnormal{F}_\textnormal{p}=\sqrt{\delta \overline{\textnormal{PSD}}^2}=\frac{\delta \textnormal{P}_\textnormal{tot}}{\textnormal{N}_\textnormal{bin}}\; .} 
\end{equation}}

\bl{We used the central limit theorem and re-binned the spectrum by a factor of $n=50$. The total amount of power in the spectrum is}
\begin{equation}
\bl{\textnormal{P}_{tot}=\sum_{j}{\textnormal{P}_\textnormal{n,j}} \; ,}
\end{equation}
\bl{where $\textnormal{P}_\textnormal{n,j}$ follows a quasi-normal distribution with $2\textnormal{n}$ degrees of freedom. It  assumes that the signal does not change dramatically over this range of $50$ bins, which is a strong assumption for classical pulsators. 
This leads to a global uncertainty on FliPer values of }
\begin{equation}
\bl{\delta \textnormal{F}_\textnormal{p}=\frac{\sqrt{\sum_{j} \left( 2\times \frac{\textnormal{P}_\textnormal{50,j}}{2 n} \times \sqrt{n} \right)^2 }}{\textnormal{N}_\textnormal{bin}} \; .}
\end{equation}
 \\
 
 \bl{The effective temperature values for the \textit{Kepler} set are taken directly from the \cite{2017ApJS..229...30M} catalogue. 
 As long as no spectroscopic follow-up surveys are available, only the effective temperature coming from the TIC will be available for most TESS data. Large uncertainties are expected because on average, $\overline{\delta \textsl{T}_\textnormal{eff}} \sim 170$~K, according to the first sector data. To be representative of future TESS data, we decided to use $\delta \textsl{T}_\textnormal{eff}=170$~K instead of the uncertainties from the \cite{2017ApJS..229...30M} catalogue for the whole \textit{Kepler} test sample.}\\
 
\bl{
We then included the effect of these errors on the different parameters during the testing of the algorithm. We performed a Monte Carlo simulation by generating for each star in our test sample $100$ artificial parameter sets from their corresponding normal distributions. We computed for each $\textnormal{X}$ parameter (F$_{{\textnormal{p}_{0.7}}}$, F$_{{\textnormal{p}_{7}}}$, F$_{{\textnormal{p}_{20}}}$, F$_{{\textnormal{p}_{50}}}$, and $\textsl{T}_{\textnormal{eff}}$) $100$ new values $\textnormal{X}_{0\le i \le 100}$ following }

\begin{equation}
\bl{\textnormal{X}_{0\le i \le 100}=\textnormal{X}+\delta \textnormal{X} \times \mathcal{G}_{0\le i \le 100} \; ,
\label{variable}}
\end{equation}
\bl{
with $\mathcal{G}_{0\le i \le 100}$ being $100$ random values following the standard distribution. Each new group of $\textnormal{X}_{i}$ parameters describes a new star to test the algorithm.\\}

\bl{We continued to train the algorithm using the original \textit{Kepler} training set. The new test set now contained a hundred times more stars than the original test set in order to include the effect of uncertainties on the input parameters. We are able to classify these new stars with a $99\%$ accuracy. We thus conclude that there is no effect of uncertainties of the chosen input parameters for the classification of stars because the classes are well separated in the $\log$(\textsl{T}$_{\rm eff}$) versus $\log(\textnormal{F}_{\textnormal{p},0.7})$ (see Fig.~\ref{Fig1}). In particular, large uncertainties on \textsl{T}$_{\textnormal{eff}}$ that are representative of future TIC effective temperatures do not perturb the pulsator classification. }

\section{\re{Conclusion}}

The study on \emph{Kepler} data confirms the results obtained by using TESS-simulated data. As expected, FliPer$_{\textnormal{Class}}$ is a great method to recognise solar-like stars based on the shape of their PSD. Using $\textnormal{F}_{\textnormal{p},i}$ ($i \in [0.7, 7, 20, 50]$ $\mu$Hz) along with \textsl{T}$_{\rm eff}$ as input parameters in a random forest algorithm, we classified more than $98\%$ of TESS-simulated and almost all the \emph{Kepler} solar-like pulsators within the test set amongst other pulsators. {We plan to improve the $\textnormal{F}_{\textnormal{p},i}$ calculation (especially for stars observed by TESS with a low signal-to-noise ratio) by empirically calibrating the photon noise as a function of the TESS magnitude of the star \citep[similar to the study by][for the \emph{Kepler} data]{2010ApJ...713L.120J} instead of measuring the power at high frequencies, which can be biased by astrophysical signal.
By comparing the results on noisy data with previous results obtained using clean simulated data \citep{arXiv:1811.12140}, we note that the performance of FliPer$_{\textnormal{Class}}$ is only slightly diminished by photometric noise. This is also auspicious for the applicability of the method to real TESS data.} This study will help the massive seismic analysis of TESS solar-like stars with global seismic pipelines by providing a list of stars that are predicted to be solar-like stars.\\

FliPer$_{\rm Class}$ gives a high weight to seismology through the use of the F$_{\textnormal{p},i}$ parameters. We chose not to incorporate any \textsl{Gaia} parameters in the FliPer$_{\rm Class}$ to remain as general as possible. For example, for faint stars such as the \emph{Kepler} RGs at the deep end of the Milky Way \citep{2016ApJ...833..294M} or for polluted systems, \textsl{Gaia} luminosities could have large uncertainties or might even be biased. Hence, seismic parameters coupled to effective temperature could be a better choice, as shown by \cite{2017ApJ...844..102H}. Therefore, the FliPer$_{\rm Class}$ as defined here could be complemented by any additional precise astrometric, photometric, or spectroscopic parameters, which could then be applied to any observations from \emph{Kepler}, K2, or TESS missions.\\

FliPer parameters are integrated as features in the TASOC/T'DA random forest classifier that will be used to automatically classify all TESS targets. This enlarged random forest is itself part of a larger classifier that includes convolutional neural networks \citep{2018MNRAS.tmp..471H}, clustering, etc. The pipeline (Tkatchenko et al., \emph{in prep}) is currently being built to be efficient in classifying all types of pulsators, and should demonstrate a high level of performance even for stars with complicated pulsation patterns.\\

\begin{acknowledgements}
We thank the enitre T'DA team for useful comments and discussions, in particular 
Andrew Tkachenko. We also acknowledge Marc Hon, Keaton Bell, and James Kuszlewicz for useful comments on the manuscript. L.B. and R.A.G. acknowledge the support from PLATO and GOLF CNES grants. S.M. acknowledges support by the Ramon y Cajal fellowship number RYC-2015-17697. O.J.H. and B.M.R. acknowledge the support of the UK Science and Technology Facilities Council (STFC). M.N.L. acknowledges the support of the ESA PRODEX programme (PEA 4000119301). Funding for the Stellar Astrophysics Centre is provided by the Danish National Research Foundation (Grant DNRF106).
\end{acknowledgements}

\bibliographystyle{aa}  
\bibliography{BIBLIO}
\begin{appendix}
\section{Random forest classifier \label{random}}

\subsection{Supervised machine-learning}
\ree{Random forests are supervised machine-learning (ML) algorithms, which learn how to predict an output variable ($\textnormal{Y}_{\textnormal{predicted}}$) from some training data (X) for which the corresponding result ($\textnormal{Y}_{\textnormal{known}}$) is already known. They learn a mapping function \textit{f} from the input(s) to the output:
\begin{equation}
\textnormal{Y}_{\textnormal{predicted}}=\textit{f}(X)
\end{equation}
The algorithm iteratively makes predictions ($\textnormal{Y}_{\textnormal{predicted}}$) on the training data (X). They are corrected to achieve a maximum level of performance by comparing with the $\textnormal{Y}_{\textnormal{known}}$ classes. The out-of-bag (OOB) error evaluates at each step the performance of the algorithm. We use a surpervised ML algorithm for our study because we have input variables X (which are F$_{{\textnormal{p}_{0.7}}}$, F$_{{\textnormal{p}_{7}}}$, F$_{{\textnormal{p}_{20}}}$, F$_{{\textnormal{p}_{50}}}$, and \textsl{T}$_{\rm eff}$) and an output class $\textnormal{Y}_\textnormal{known}$ (representing the type of pulsator).}

\subsection{Classification trees}
\ree{
The clasification-tree method is part of the Classification and Regression Trees (CART) introduced by \cite{Breiman2001}. A decision-tree algorithm constructs a binary tree during the training, with each node representing a split point on an input variable (X) (numerical value for regression algorithms, or class name for classification algorithms). The leaf nodes of the tree contain the possible output classes ($\textnormal{Y}_\textnormal{predicted}$).}

\ree{The tree is built such that a cost function is minimized at each node. Equation~\ref{cost} is the cost function used for the classifier, with N$_{\textnormal{classes}}$ the number of classes and $p_k$ the number of training instances with class $k$ at the node of interest, }

\ree{
\begin{equation}
\label{cost}
\textnormal{G}=\sum_{k=1}^{\textnormal{N}_{\textnormal{classes}}}p_k \times(1-p_k).
\end{equation}
When the tree is built on the training sample, it is used to evaluate $\textnormal{Y}_\textnormal{predicted}$ for new $\textnormal{X}_\textnormal{new}$ data.}

\ree{\subsection{Ensemble method Random Forest classifier}}
\ree{An ensemble method combines the prediction from multiple ML algorithms. It aims at making even more accurate predictions than any individual model. The Random Forest classifier is an ensemble method that combines classification trees. It consists of the following steps:
\begin{itemize}
\item{Creating many subsamples of the training sample.}
\item{Training a classification tree on each subsample, keeping a low number of features that can be looked at for each split point. It aims at decreasing the correlation between the different trees. For classification algorithms, the maximum number of features searched for at each split point is usually $m=\sqrt{p}$, where $p$ is the number of input (X) variables.}
\item{Calculating the dominant class from each model for the new test sample: this predicted class is used as the output variable ($\textnormal{Y}_\textnormal{predicted}$).}
\end{itemize}}

\end{appendix}
\end{document}